\documentclass{emulateapj}

\begin{document}

\title{New Ultraviolet Extinction Curves for Interstellar Dust in
M31\altaffilmark{*}}

\author{Geoffrey C. Clayton\altaffilmark{1},
Karl~D.~Gordon\altaffilmark{2,3}, Luciana C. Bianchi\altaffilmark{4},
Derck L. Massa\altaffilmark{5}, Edward L. Fitzpatrick\altaffilmark{6},
R. C. Bohlin\altaffilmark{2}, and Michael J. Wolff\altaffilmark{5}}

\altaffiltext{1}{Dept.\ of Physics \& Astronomy, Louisiana State
University, Baton Rouge, LA 70803; gclayton@fenway.phys.lsu.edu}

\altaffiltext{2}{Space Telescope Science Institute; 3700 San Martin
Drive, Baltimore, MD 21218; bohlin, kgordon@stsci.edu}

\altaffiltext{3}{Sterrenkundig Observatorium, Universiteit Gent,
              Gent, Belgium}

\altaffiltext{4}{Department of Physics and Astronomy, The Johns
Hopkins University, 3400 N. Charles St., Baltimore, MD 21218;
bianchi@jhu.edu }

\altaffiltext{5}{Space Science Institute, 4750 Walnut St. Suite 205,
Boulder, CO 80301, USA; mjwolff@spacescience.org}

\altaffiltext{6}{Department of Astronomy and Astrophysics, Villanova
University, 800 Lancaster Avenue, Villanova, PA 19085, USA;
edward.fitzpatrick@villanova.edu}

\altaffiltext{*} {Based on observations made with the NASA/ESA {\it
Hubble Space Telescope}, obtained by the Space Telescope Science
Institute, and from the data archive at STScI\null. STScI is operated
by the Association of Universities for Research in Astronomy, Inc.,
under NASA contract NAS5-26555.}

\begin{abstract}
New low-resolution UV spectra of a sample of reddened OB stars in M31
were obtained with {\it HST}/STIS to study the wavelength dependence
of interstellar extinction and the nature of the underlying dust grain
populations. Extinction curves were constructed for four reddened
sightlines in M31 paired with closely matching stellar atmosphere
models. The new curves have a much higher S/N than previous
studies. Direct measurements of N(H I) were made using the Ly$\alpha$
absorption lines enabling gas-to-dust ratios to be calculated. The
sightlines have a range in galactocentric distance of 5 to 14 kpc and
represent dust from regions of different metallicities and gas-to-dust
ratios. The metallicities sampled range from Solar to 1.5 Solar. The
measured curves show similarity to those seen in the Milky Way and the
Large Magellanic Cloud. The Maximum Entropy Method was used to
investigate the dust composition and size distribution for the
sightlines observed in this program finding that the extinction curves
can be produced with the available carbon and silicon abundances if
the metallicity is super-Solar.
\end{abstract}

\keywords{dust, extinction --- ISM: abundances --- ISM: molecules ---
          ultraviolet: ISM}

\section{Introduction}

\citet[][CCM]{1989ApJ...345..245C} found that the wavelength
dependence of extinction in the Galaxy was a function of one
parameter, R$_V$ (=A$_V$/E(B-V)), the ratio of total to selective
extinction. This work was extended to over 400 sightlines in the
Galaxy of which only 4 differed significantly from CCM
\citep{2004ApJ...616..912V}. However, it has become increasingly
apparent that ``standard'' Milky Way (MW) type dust extinction does
not generally apply to interstellar dust in other galaxies. It has
been known for a long time that the UV extinction properties in the
Magellanic Clouds are different from the MW and from each other.  In
particular, sightlines near 30 Dor in the Large Magellanic Cloud (LMC)
\citep{1985ApJ...288..558C,1985ApJ...299..219F} and in the
star-forming Bar of the SMC \citep{1984A&A...132..389P} show very
un-MW extinction curves, especially in their weak 2175 \AA\ bumps and
steep far-UV rises.  The variations seen in dust properties extend
beyond that of the extinction curve shape, including e.g., the
gas-to-dust ratio, indicating that there may be a wide range of
underlying grain populations.  In addition, many starburst galaxies
seem to manifest dust properties similar to those associated with the
Small Magellanic Cloud (SMC) \citep{1998ApJ...500..816G}.

The CCM study only probed a limited set of dust environments in the
MW. All of the sightlines were within 1 kpc of the Sun, so it is not
surprising that extinction in other galaxies has different properties
than the MW. The observed variations from galaxy to galaxy are an
indication of the complexity of dust properties and demonstrate the
need for direct studies of dust for environments more representative
of the current sample of extragalactic SEDs
\citep[e.g.,][]{1994ApJ...429..582C,1997ApJ...487..625G}.

The wavelength dependence of UV extinction due to dust can only be
directly measured along sightlines where the spectra of individual
stars can be obtained. Otherwise, the dust extinction characteristics
must be inferred through radiative transfer modeling
\citep{1994ApJ...429..582C,1997ApJ...487..625G}.  This technique is
therefore limited to galaxies in the Local Group. Almost all of the
studies of UV extinction have been limited to stars in three galaxies,
the MW
\citep{1984ApJ...279..698W,1988A&AS...73..195A,1990ApJS...72..163F,2000ApJS..129..147C,2004ApJ...616..912V,2007ApJ...663..320F},
the LMC
\citep{1985ApJ...288..558C,1985ApJ...299..219F,1999ApJ...515..128M,2003ApJ...594..279G,2014MNRAS.445...93D}
and the SMC
\citep{1982A&A...113L..15L,1984A&A...132..389P,1998ApJ...500..816G,2003ApJ...594..279G,2012A&A...541A..54M,gordon2015}.

Beyond the MW and the Magellanic Clouds, there isn't much information
on the extinction properties in other Local Group galaxies.  Radiative
transfer modeling of the emission from the starburst nucleus of M33
indicates the presence of a strong 2175 \AA\ bump unlike other
starbursts which show more SMC-like extinction having little or no
evidence for a bump \citep{1999ApJ...519..165G}.  A recent {\it Swift}
and {\it HST} photometric study of dust clumps in the bulge of M31
suggests that the UV extinction is generally steeper than the CCM,
R$_V$=3.1 curve \citep{2014ApJ...785..136D}.  Analysis of interstellar
polarization along four sightlines in M31 suggests significant
differences in the size distribution of silicate grains compared with
those studied elsewhere in the Local Group
\citep{2004AJ....127.3382C}.

\section{Dust and Gas in M31}

Almost 20 years ago, \citet{1996ApJ...471..203B} did a ``pilot study''
using {\it HST}/FOS of UV extinction toward a very small sample of OB
stars in M31. They derived an average M31 extinction curve from only
three sightlines that has an overall wavelength dependence similar to
that of the average Galactic extinction curve, but potentially
possessing a weaker 2175 \AA\ bump. While the extinction curves
calculated from these data provided a proof-of-concept, the study
suffered several drawbacks that made interpretation of the extinction
difficult: the UV spectra had very low S/N and the stars observed were
only lightly reddened.

The metallicity of M31 has been determined to be about two times Solar
by \citet{1994ApJ...420...87Z} based on H II regions. Studies of some
A and B supergiants in M31 suggested that the metallicity was close to
Solar \citep{2000ApJ...541..610V,2001MNRAS.325..257S} but a large,
recent study by \citet{2012ApJ...758..133S} has reconfirmed the
super-Solar abundances. 

In this paper, we present new high S/N {\it HST}/STIS spectra of
a small number of significantly reddened stars in M31 and
construct extinction curves for those sightlines to study the nature
of its interstellar dust.

\section{Observations}

\begin{deluxetable*}{llllllll}
\tablecaption{New and Archival {\it HST}/STIS Observations in M31}
\tablehead{\colhead{Star}&\colhead{MAC86$^a$}&
           \colhead{RA}&
           \colhead{Dec}&
           \colhead{Date}&
            \colhead{Exp.\ (s)}&
            \colhead{Dataset}&
             \colhead{STIS Grating} }
\startdata
J003733.35+400036.6&\nodata&00 37 33.340&+40 00 36.70&2012-10-05&2423&OBPX01010&G230L\\
&&&&2000-07-07&2323&O56R12010&G140L$^b$\\
&&&&2000-07-07&2323&O56R12020&G140L$^b$\\
J003944.71+402056.2&\nodata&00 39 44.710&+40 20 56.20&2013-01-27&2423&OBPX02010&G230L\\
&&&&2013-01-27 &2908&OBPX02020&G140L$^c$\\
&&&&2013-07-13 &2494&OBPX52010&G140L\\
J003958.22+402329.0&\nodata&00 39 58.220&+40 23 29.00&2013-02-03 &2423&OBPX03010&G230L\\
&&&&2013-02-03 &3077&OBPX03020&G140L\\
J004029.71+404429.8$^d$&OB 78-231&00 40 29.700&+40 44 28.40&2012-10-04 &2423&OBPX04010&G230L\\
 &&&&2004-01-22&1820&O8MG01010&G140L$^e$\\
&&&&2004-01-22&2840&O8MG01020&G140L$^e$\\
&&&&2004-01-22&2800&O8MG01030&G140L$^e$\\
&&&&2004-01-22&2800&O8MG01040&G140L$^e$\\
&&&&2004-01-22&2800&O8MG01050&G140L$^e$\\
J004030.94+404246.9&OB 78-347&00 40 30.940&+40 42 46.90&2013-02-11 &2423&OBPX05010&G230L\\
&&&&2013-02-11 &3077&OBPX05020&G140L\\
J004031.52+404501.9&OB 78-376&00 40 31.520&+40 45 01.90&2012-12-19 &2423&OBPX06010&G230L\\
&&&&2012-12-19 &3077&OBPX06020&G140L\\
J004034.61+404326.1&OB 78-550&00 40 34.610&+40 43 26.10&2013-02-09 &2423&OBPX07010&G230L\\
&&&&2013-02-09 &3077&OBPX07020&G140L\\
J004037.92+404333.3&\nodata&00 40 37.920&+40 43 33.30&2013-02-10 &2423&OBPX08010&G230L\\
&&&&2013-02-10 &3077&OBPX08020&G140L\\
J004412.17+413324.2&\nodata&00 44 12.170&+41 33 24.20&2012-06-28 &2423&OBPX09010&G230L\\
&&&&2012-06-28 &3077&OBPX09020&G140L\\
J004412.97+413328.8&OB 10-150&00 44 12.970&+41 33 28.80&2012-10-03 &2423&OBPX10010&G230L\\
&&&&2003-09-27 &2200&O8MG07010&G140L$^e$\\
&&&&2003-09-27 &2840&O8MG07020&G140L$^e$\\
&&&&2003-09-27 &2800&O8MG07030&G140L$^e$\\
&&&&2003-09-27 &2800&O8MG07040&G140L$^e$\\
&&&&2003-09-27 &2800&O8MG07050&G140L$^e$\\
&&&&2003-09-27 &2840&O8MG08010&G140L$^e$\\
&&&&2003-09-27 &2840&O8MG08020&G140L$^e$\\
&&&&2003-09-27 &2800&O8MG08030&G140L$^e$\\
&&&&2003-09-27 &2800&O8MG08040&G140L$^e$\\
&&&&2003-09-27 &2800&O8MG08050&G140L$^e$\\
J004515.27+413747.9$^e$&OB 48-444&00 45 15.270&+41 37 47.90&2012-12-16 &2423&OBPX11010&G230L\\
&&&&2012-12-16 &0&OBPX11020&G140L$^f$\\
&&&&2013-01-21 &2494&OBPX13010&G140L
\enddata
\tablenotetext{a}{\citet{2006AJ....131.2478M}}
\tablenotetext{b}{\citet{2002ApJ...580..213B}, 52\arcsec x 0\farcs2 aperture}
\tablenotetext{c}{G140L failed. Re-observed on 2013-07-13}
\tablenotetext{d}{Target in \citet{1996ApJ...471..203B}}
\tablenotetext{e}{Data obtained for GO 9794.}
\tablenotetext{f}{G140L failed. Re-observed on 2013-01-21}
\label{tab_stis_obs}
\end{deluxetable*}

\begin{deluxetable*}{lllllll}
\tablecaption{Stellar Parameters}
\tablehead{\colhead{Star}&\colhead{V}&
           \colhead{B-V}&
           \colhead{Sp.T.}&
            \colhead{E(B-V)$^a$}&
             \colhead{Ref$^b$} }
\startdata
J003733.35+400036.6&	18.16&	-0.21&	B2 Ia		&-0.04			&2\\	
J003944.71+402056.2&	18.2&	0.15&	O9.7 Ib	&		0.42&1\\		
J003958.22+402329.0&	18.97&	0.09&	B0.7 Ia	&		0.30&1\\
J004029.71+404429.8&	18.56&	-0.23&	O7-7.5 Iaf	&		0.05&1,3\\
J004030.94+404246.9&	18.87&	-0.15&	O9.5 Ib&0.12&1\\
J004031.52+404501.9&	18.92&	-0.15&	B0.5 Ia& 0.07& 1\\
J004034.61+404326.1&	18.67&	0.15&	B1 Ia	&	0.34&1\\
J004037.92+404333.3&	18.66&	0.06&	B1.5 Ia&			0.23&1\\
J004412.17+413324.2	&17.33	&0.34&	B2.5 Ia	&		0.49&1\\							
J004412.97+413328.8&	19.18&	-0.04	& O8.5 Ia(f)	&			0.24&3\\
J004515.27+413747.9&	19.10	&-0.02&	O8 I			&	0.26&3
\enddata
\tablenotetext{a}{Total E(B-V) including MW foreground assuming the measured spectral type and intrinsic colors \citep{1970A&A.....4..234F}.}
\tablenotetext{b}{Spectral types are from: 1) \citet{2011ApJ...726...39C}, 2) \citet{2002ApJ...580..213B}, 3) \citet{1995AJ....110.2715M}.}
\label{tab_stell_param}
\end{deluxetable*}

Eleven early-type M31 supergiants were observed by {\it HST}/STIS
using the G140L and G230L gratings. The sample was selected from stars
known to be members of M31 for which spectral types are available
\citep{1995AJ....110.2715M, 2002ApJ...580..213B,
2011ApJ...726...39C}. Two of the G140L observations, marked in
Table~\ref{tab_stis_obs}, failed and were redone. Also, three stars in
the sample already had 
existing G140L observations. These were retrieved from the {\it
Mikulski Archive for Space Telescopes} (MAST) and were used in this
study. The new and archival observational data are summarized in
Table~\ref{tab_stis_obs}. Table~\ref{tab_stell_param} lists the
stellar parameters for the 
sample. The new UV 
spectra have been combined with existing ground-based UBVRI photometry
\citep{2006AJ....131.2478M} to create an SED for each star.

The sample of stars, for which new and archival {\it HST}/STIS
spectra have been obtained, consists of seven significantly reddened
stars (E(B-V)= 0.3-0.5) and four lightly reddened stars
(E(B-V)$\sim$0.1) of similar spectral types. These stars are all
members of M31 \citep{2006AJ....131.2478M}. The sample is limited to
spectral types ranging from O7 to B2.5 supergiants
\citep{1995AJ....110.2715M, 2002ApJ...580..213B, 2011ApJ...726...39C}.

Our most reddened star, J004412.17+413324.2, is very bright (V=17.33
and B-V = 0.34 \citep{2006AJ....131.2478M}).  If it is an early B
supergiant then the expected M$_V$ $\sim$ -6.5. With a calculated
reddening E(B-V)=0.49 and assuming a distance modulus to M31 of 24.4
\citep[e.g.,][]{2012ApJ...745..156R}, then M$_V$ = 17.34 - 3.1 x
(0.49) -24.4 = -8.6 mag. This is consistent with a bright blue
supergiant and it has been confirmed as a member of M31
\citep{2006AJ....131.2478M}. \citet{2003AJ....126..175B} found
J004412.17+413324.2 to be a variable star.

Only one reddened star in our sample, J003944.71+402056.2, has
near-IR photometry (J=17.83, K=17.68) \citep{2014AJ....147..109S}. For
this star, E(V-K) = 1.25 assuming it is an O9.7 Ib star
\citep{2000asqu.book.....C}. Then, A(V) can be estimated by
E(V-K)$\times$1.1 = 1.37 mag \citep{1978A&A....66...57W}, and R$_V$=
A(V)/E(B-V) = 1.37/0.42 = 3.3.

\section{Extinction Curves}

Traditionally, extinction curves are calculated using the standard
pair method \citep{1983ApJ...266..662M,1992AJ....104.1916C}, which
requires a reddened star and a lightly reddened comparison star having
the same or similar spectral type.  However, finding a a good spectral
match is not easy especially when a limited sample of lightly reddened
stars is available, so stellar atmosphere models are now being
extensively used as pair stars. The use of spectra from stellar
atmosphere models as comparison ``stars'' is at least as accurate as
using actual stellar spectra \citep{2005AJ....130.1127F}.  Extinction
curves were attempted using the reddened and lightly reddened stars in
the sample but the spectral type matches were not
satisfactory. Extinction curve matches were made with stellar
atmosphere models with better success.

\begin{deluxetable}{cccc}
\tablecaption{Model Parameters}
\tablehead{   
 \colhead{Parameter} & \colhead{Description} &
    \colhead{Min} &
    \colhead{Max} }
\startdata  
\multicolumn{4}{c}{M31 Components} \\ \hline
$\log(T_\mathrm{eff})$ & effective temperature & 4.18 & 4.74 \\
$\log(g)$  & surface gravity & 1.75 & 4.75 \\
$\log(Z)$  & metallicity & -0.3 & 0.3 \\ 
$A(V)$  & V band extinction & 0.0 & 4.0 \\
$R(V)$ & $A(V)/E(B-V)$ & 1.0 & 7.0 \\
$c_2$ & UV slope & -0.5 & 1.5 \\
$c_3$ & 2175~\AA\ bump height & 0.0 & 6.0 \\
$c_4$ & FUV curvature & -0.2 & 2.0 \\
$x_0$ & 2175~\AA\ bump centroid & 4.55 & 4.65 \\
$\gamma$ & 2175~\AA\ bump width & 0. & 2.5 \\ 
$\log(HI)$ & M31 HI column & 19.0 & 24.0 \\ 
\multicolumn{4}{c}{MW Components} \\ \hline
$\log_\mathrm{MW}(HI)$ & MW HI column & 18.0 & 22.0 \\
$E(B-V)_{MW}$ & MW dust column & \multicolumn{2}{c}{0.06} \\
$R(V)_{MW}$ & MW $A(V)/E(B-V)$ & \multicolumn{2}{c}{3.1} \\
$v(MW)$  & velocity [km s$^{-1}$] & \multicolumn{2}{c}{0} 
\enddata
\label{tab_fit_params}
\end{deluxetable}

The extinction curves were calculated by forward modeling of the
spectrum of each star to determine the correct model atmosphere to use
as the unreddened comparison star.  The model we adopt for the M31
stars is a combination of a TLusty stellar model atmosphere
\citep{Lanz03,2007ApJS..169...83L} extinguished by an $R(V)$-dependent
extinction curve \citep{Fitzpatrick99review} in the optical and NIR,
combined with a \citet[][FM]{1990ApJS...72..163F} parametrization in
the UV, updated by \citet{2007ApJ...663..320F}.  The assumed model fit
parameters are given in Table~\ref{tab_fit_params}.  Some of the model
parameters are fixed at single values as our observations are not
sensitive to these parameters.
The M31
components of the model have radial velocities calculated using the method of \citet{1970ApJ...159..379R} and \citet{2009ApJ...703..441D}.
These velocities are listed in Table 4. 
 The MW components of the model
have a fixed radial velocity of 0~km~s$^{-1}$.

We fit this model 
to the observed data using the EMCEE fitting code
\citep{2013PASP..125..306F}.  The observed and model spectra were
normalized by the average of the optical photometry prior to fitting.
We imposed flat priors on most of the fit parameters with the min and
max values given in Table~\ref{tab_fit_params}.  The min and max
values were set to reasonable limits on fit parameters and were
generally based on expected ranges from MW measurements of extinction
curves \citep{2004ApJ...616..912V}.  The min and max values for
$\log(Z)$ were set to be between -0.3 and 0.3 $\times$ solar metallicity
as a reasonable range given the galactocentric distances of the reddened stars in our sample.
For the stellar parameters, $\log(T_\mathrm{eff})$, $\log(g)$, and
$\log(Z)$, the base priors are given by the allowed model
space as defined by the TLusty stellar atmosphere grid.  For
$\log(T_\mathrm{eff})$, we add an
additional multiplicative Gaussian prior based on the
literature spectral type (Table~\ref{tab_stell_param}) and assuming an
uncertainty of one subclass in spectral type.  We use the
\citet{2008flhs.book.....C} calibration of spectral type to
$\log(T_\mathrm{eff})$.  The
results of the fitting have been used to calculate uncertainties on all the fit
parameters.  These uncertainties include a full accounting of the
sources of noise, including spectral mismatch illustrating one of the
strengths of the method.

\begin{deluxetable*}{lrrrrr}
\tablecaption{Stellar Parameter Results}
\tablehead{ & \multicolumn{2}{c}{Prior} & \multicolumn{3}{c}{Fit Parameters} \\
           \colhead{Star} &
           \colhead{$\log(T_{\mathrm{eff}}$)} &
           \colhead{RV} &
           \colhead{$\log(T_{\mathrm{eff}}$)} &
           \colhead{$\log(g)$} &
           \colhead{$\log(Z)$}\\
           \colhead{}&\colhead{}&\colhead{(km s$^{-1}$)}
}
\startdata
J003944.71+402056.2 & $4.47 \pm 0.03$ & -511 & $4.48^{+0.03}_{-0.02}$ & $2.79^{+0.21}_{-0.22}$ & $0.03^{+0.16}_{-0.17}$ \\
J003958.22+402329.0 & $4.37 \pm 0.04$ & -505 & $4.35^{+0.04}_{-0.03}$ & $2.08^{+0.52}_{-0.24}$ & $0.02^{+0.19}_{-0.19}$ \\
J004034.61+404326.1 & $4.33 \pm 0.02$ & -536 & $4.35^{+0.02}_{-0.02}$ & $2.53^{+0.47}_{-0.23}$ & $0.16^{+0.09}_{-0.17}$ \\
J004412.17+413324.2 & $4.23 \pm 0.04$ & -76  & $4.24^{+0.04}_{-0.04}$ & $2.32^{+1.11}_{-0.39}$ & $0.03^{+0.18}_{-0.21}$ 
\enddata
\label{tab_fit_results_stellar}
\end{deluxetable*}

\begin{deluxetable*}{llllllll}
\tablecaption{Dust FM Parameter Results$^a$}
\tablehead{\colhead{Star} &
           \colhead{$E(B-V)$}&
           \colhead{$c_1^b$}&
            \colhead{$c_2$}&
             \colhead{$c_3$}&
              \colhead{$c_4$}&
               \colhead{$x_0$}&
                \colhead{$\gamma$}}
\startdata
J003944.71+402056.2 & $0.35^{+0.04}_{-0.04}$ & $0.24^{+0.28}_{-0.31}$ & $0.65^{+0.11}_{-0.10}$ & $3.58^{+1.21}_{-1.01}$ & 
   $0.41^{+0.08}_{-0.07}$ & $4.57^{+0.03}_{-0.02}$ & $1.07^{+0.13}_{-0.14}$ \\
   &0.37&0.33&0.62&3.05&0.38&4.57&0.99\\
J003958.22+402329.0 & $0.25^{+0.04}_{-0.03}$ & $-0.30^{+0.54}_{-0.68}$ & $0.84^{+0.24}_{-0.19}$ & $3.63^{+1.55}_{-1.72}$ & 
   $-0.02^{+0.09}_{-0.09}$ & $4.62^{+0.02}_{-0.03}$ & $1.13^{+0.18}_{-0.24}$ \\
      &0.24&-0.41&0.88&3.95&-0.01&4.65&1.15\\
J004034.61+404326.1 & $0.26^{+0.03}_{-0.03}$ & $-0.95^{+0.45}_{-0.51}$ & $1.07^{+0.18}_{-0.16}$ & $4.27^{+1.16}_{-1.35}$ & 
   $-0.14^{+0.07}_{-0.04}$ & $4.63^{+0.01}_{-0.03}$ & $1.14^{+0.15}_{-0.16}$  \\
      &0.25&-0.86&1.04&5.16&-0.16&4.65&1.25\\
J004412.17+413324.2 & $0.38^{+0.05}_{-0.04}$ & $-1.54^{+0.51}_{-0.43}$ & $1.28^{+0.15}_{-0.18}$ & $3.93^{+1.37}_{-1.44}$ & 
   $0.04^{+0.23}_{-0.16}$ & $4.61^{+0.03}_{-0.04}$ & $1.17^{+0.22}_{-0.25}$ \\
      &0.39&-1.57&1.29&3.16&-0.17&4.64&1.06
\enddata
\tablenotetext{a}{For each star, the first line is the 50\% (median probability) fit and the second line is the best fit.}
\tablenotetext{b}{The $c_1$ parameter is not fit. It is assumed that
  $c_1$ = 2.09 - 2.84 $c_2$ \citep{2007ApJ...663..320F}.}
\label{tab_fit_results_dust}
\end{deluxetable*}

\begin{figure}[tbh]
\begin{center}
\epsscale{1.1}
\plotone{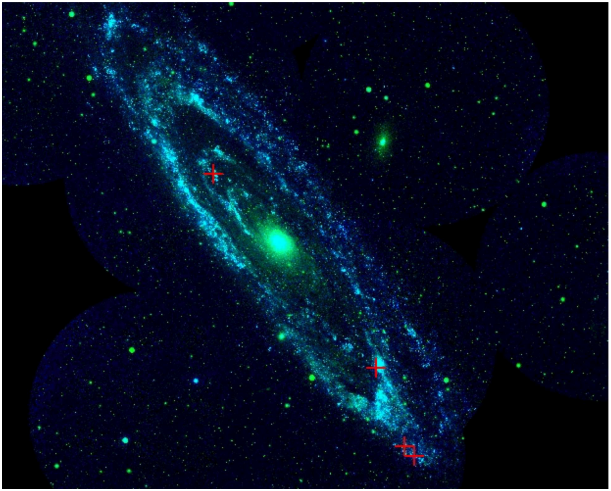}
\end{center}
\caption{Color GALEX Image of M31 (NUV green, FUV blue) with the
locations of the four reddened stars in the sample marked by red
crosses.}
\label{m31image}
\end{figure}

\begin{figure}
\begin{center}
\epsscale{1.2}
\plotone{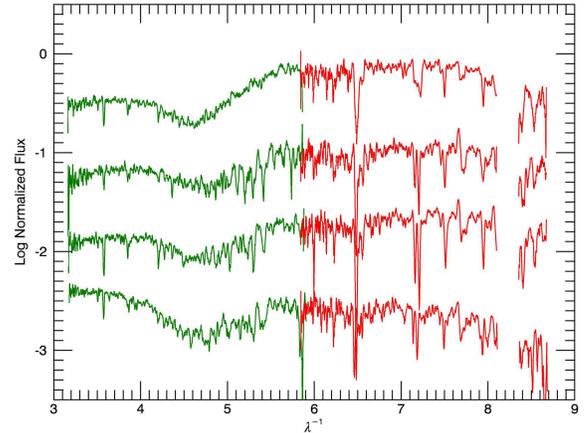}
\end{center}
\caption{{\it HST}/STIS G140L (red) and G230L (green) spectra for four
  significantly redded stars. From top to bottom, the spectra are
J003944.71+402056.2, J003958.22+402329.0, J004034.61+404326.1, and
J004412.17+413324.2. }
\label{spectra}
\end{figure}

\begin{deluxetable*}{lrrrrrr}[tbp]
\tablecaption{Gas-to-Dust Ratio Measurements}
\tablehead{ & & \multicolumn{3}{c}{Fit Results} & \multicolumn{2}{c}{21~cm Results} \\
  \colhead{Star}&
  \colhead{GC Dist.$^\tablenotemark{a}$}&
           \colhead{$E(B-V)$}&
         \colhead{$N(H I)^\tablenotemark{b}$}&
           \colhead{$N(H I)/E(B-V)^\tablenotemark{c}$}&
             \colhead{$N(H I)^\tablenotemark{d}$}&
                        \colhead{$N(H I)/E(B-V)^\tablenotemark{e}$}}
\startdata
J003944.71+402056.2 & 14.0 & $0.35^{+0.04}_{-0.04}$ & $4.2^{+1.7}_{-2.0}$ & $11.5^{+5.1}_{-5.6}$ & 4.5 & 6.4 \\
J003958.22+402329.0 & 13.3 & $0.25^{+0.04}_{-0.03}$ & $5.6^{+1.8}_{-2.0}$ & $22.0^{+8.5}_{-8.3}$ & 4.0 & 8.0 \\
J004034.61+404326.1 & 8.8  & $0.26^{+0.03}_{-0.03}$ & $6.2^{+1.6}_{-2.0}$ & $23.7^{+6.9}_{-7.8}$ & 2.3 & 4.4 \\
J004412.17+413324.2 & 5.2  & $0.38^{+0.05}_{-0.04}$ & $<2.7$              & $<7.2$               & 2.0 & 2.5 
\enddata
\tablenotetext{a}{Projected galactocentric distance from the center of M31 in kpc.}
\tablenotetext{b}{N(H I) from Ly$\alpha$ absorption. In units of 10$^{21}$ atoms cm$^{-2}$.}
\tablenotetext{c}{In units of 10$^{21}$ atoms cm$^{-2}$mag$^{-1}$.}
\tablenotetext{d}{N(H I) from 21 cm observations. In units of 10$^{21}$ atoms cm$^{-2}$ \citep{2011ApJ...726...39C}.}
\tablenotetext{e}{$0.5N(H I)/E(B-V)$. In units of 10$^{21}$ atoms cm$^{-2}$mag$^{-1}$.}
\label{tab_hi_results}
\end{deluxetable*}

\begin{figure*}
\begin{center}
\plottwo{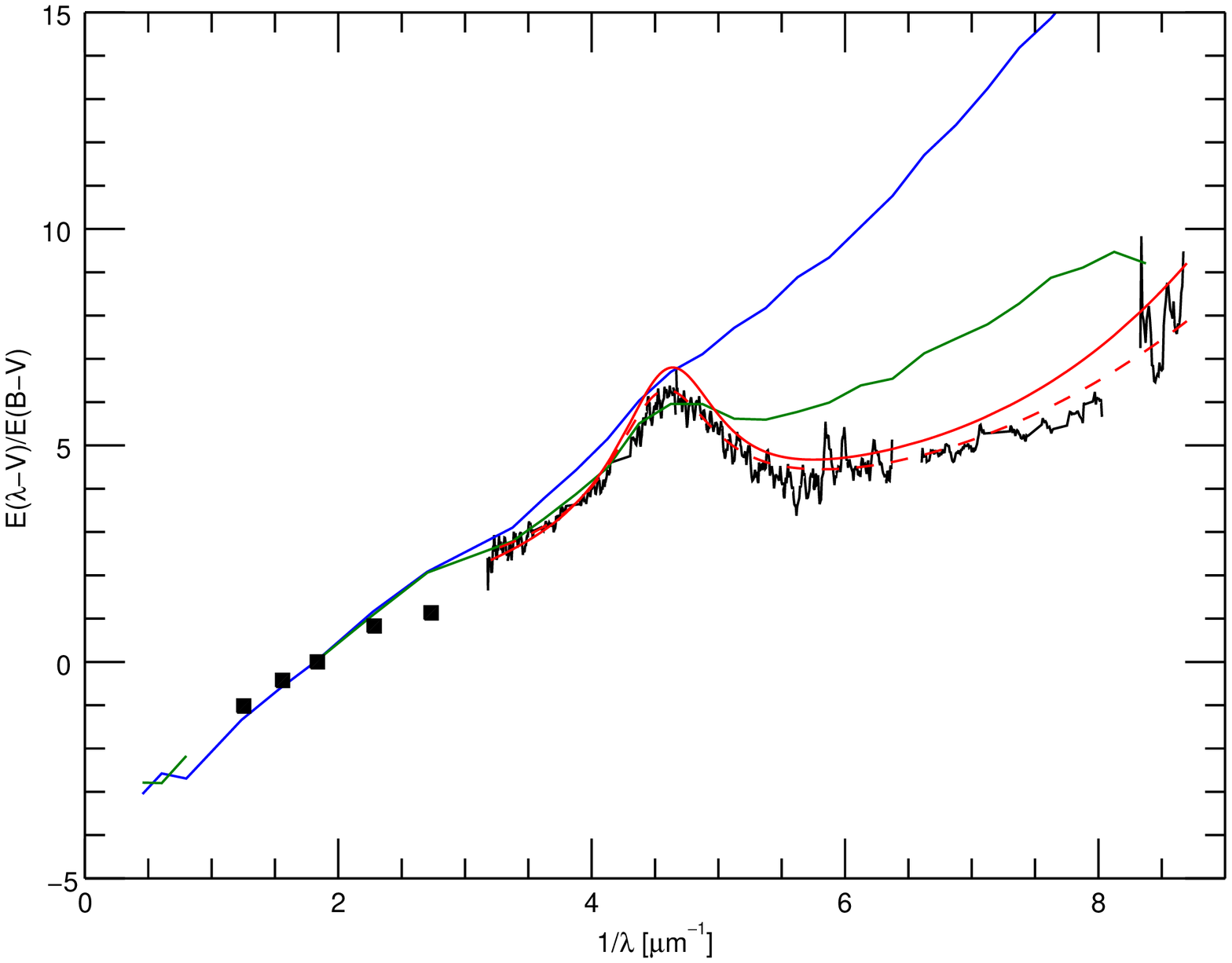}{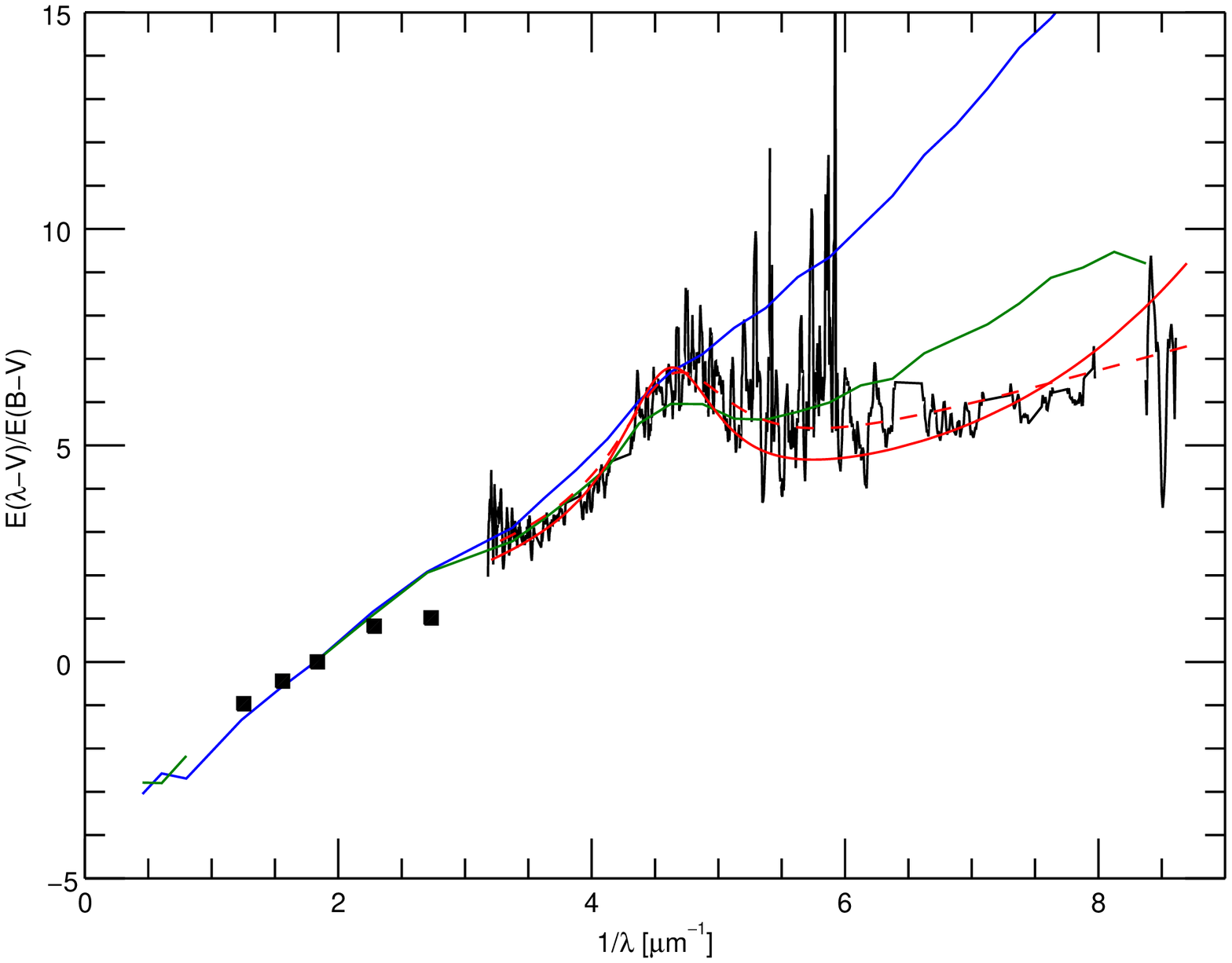} \\
\plottwo{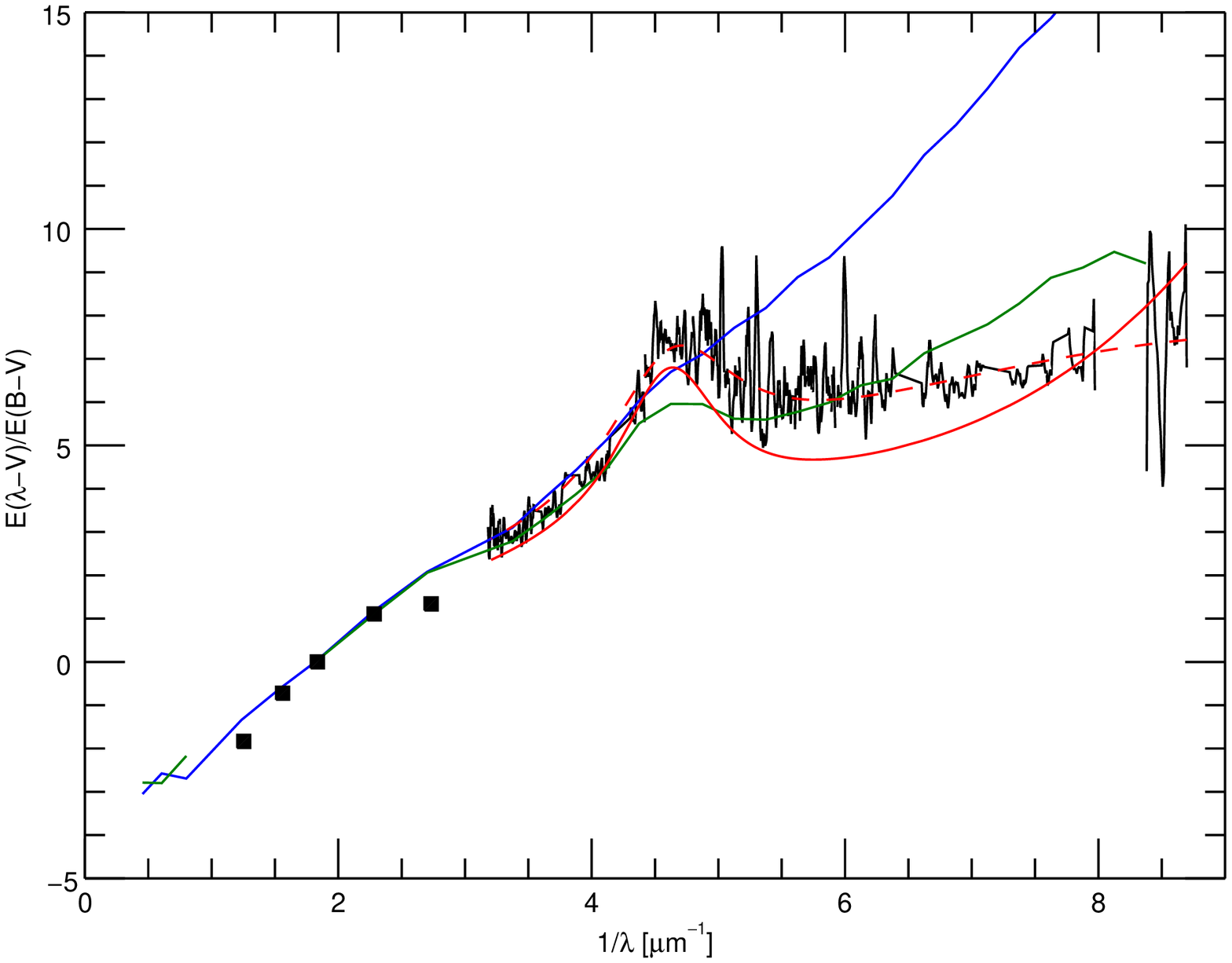}{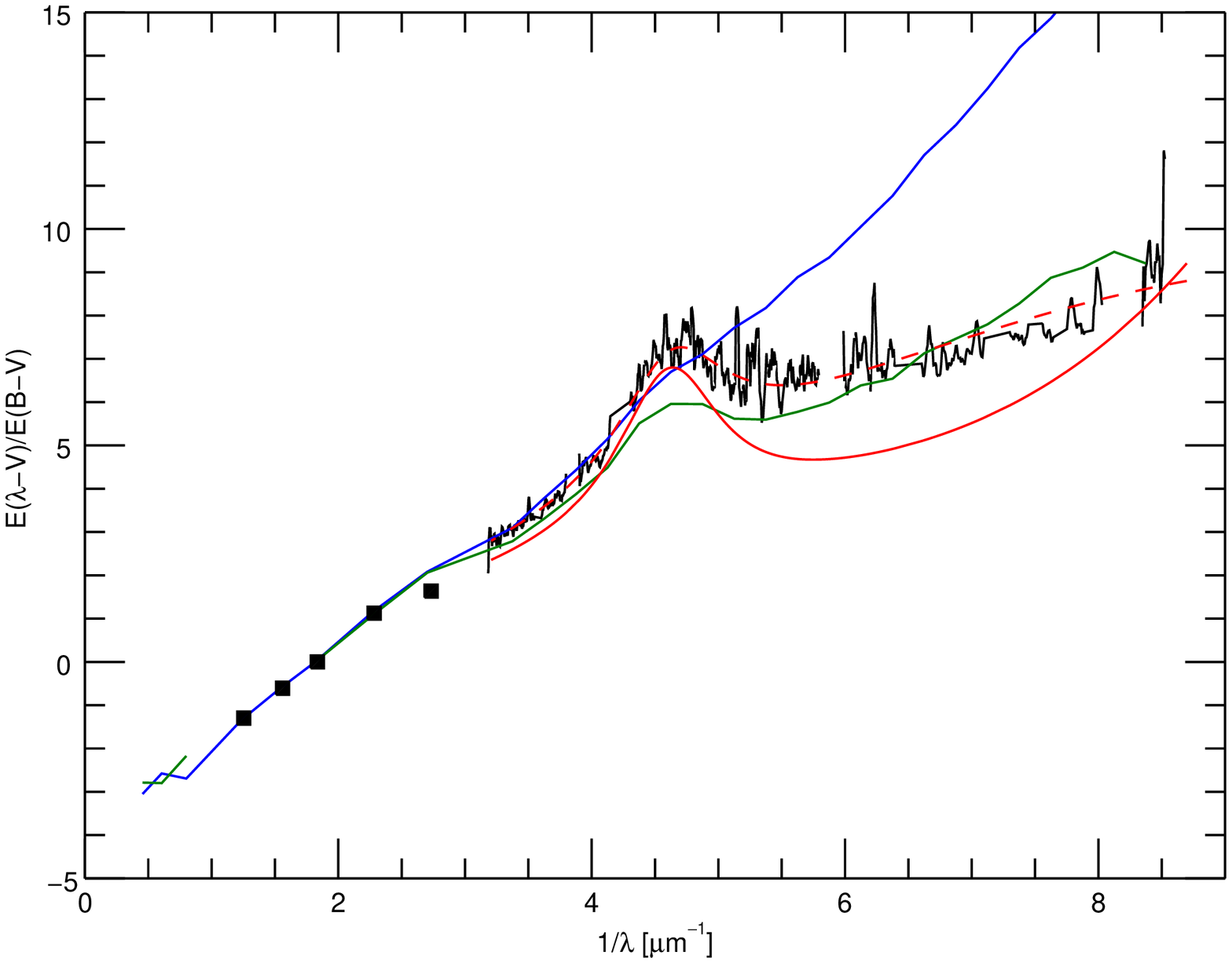}
\end{center}
\caption{Extinction curves for the target sample using the best fits
with atmosphere models. The extinction curves are J003944.71+402056.2
(upper left), J003958.22+402329.0 (upper right), J004034.61+404326.1
(lower left), and J004412.17+413324.2 (lower right). The black line is
the extinction curve using STIS spectra, and the black squares are
UBVRI. Shown for comparison are the SMC average extinction curve (blue), the
LMC (30-Dor) average extinction curve (green), and the MW CCM $R(V)=3.1$
(red). The best fit to the FM parameters, listed in Table \ref{tab_fit_results_dust}, is the red dashed line.}
\label{extcurves}
\end{figure*}

Since the stars in M31 are being paired with unreddened stellar
models, the foreground MW dust extinction must be removed explicitly.
Older estimates of the M31 foreground ($E(B-V) \sim 0.08$ mag) are
discussed in \citet{1996ApJ...471..203B}. Newer estimates including
\citet{1998ApJ...500..525S} and \citet{2011ApJ...737..103S} give an
average $E(B-V) = 0.06$ mag \citep{2011ApJ...726...39C}.  This estimate
for the MW foreground has been confirmed by a survey of M31
star-forming regions \citep{2012AJ....144..142B}.  So, a MW foreground
extinction component of $E(B-V) = 0.06$, assuming $R_V=3.1$ CCM dust, is
included as part of the fitting process described above. Three of the seven
``reddened'' stars in the sample had small E(B-V) values between 0.09
and 0.16 so these curves are not used in the analysis presented
here. The positions of the four remaining stars are plotted on a GALEX
image of M31 in Figure~\ref{m31image} and their {\it HST}/STIS spectra
are plotted in Figure~\ref{spectra}. The extinction curves for these
four significantly reddened sightlines are shown in
Figure~\ref{extcurves} with a 5-point smoothing.  The model fit
parameter results for each star are given 
in 
Tables~\ref{tab_fit_results_stellar}, \ref{tab_fit_results_dust}, and
\ref{tab_hi_results}.  
The extinction curves are fit with the \citet[][FM]{1990ApJS...72..163F} parameters. 
The fit parameters and 1$\sigma$ uncertainties
are tabulated based on the 
17\%, 50\%, and 83\% values of the marginalized 1D posterior
probability distribution functions generated from the EMCEE results.
Table \ref{tab_fit_results_dust} gives the best fit and the 50\% (median probability) fit for each extinction curve.
We give the dust column results using $E(B-V)$ as this measurement was
well behaved (small uncertainties) whereas the measurements of $A(V)$
and $R(V)$ have very large uncertainties.  The
$c_1$ parameter is not fit. We include it here assuming that $c_1 = 2.09 -
2.84c_2$ \citep{2007ApJ...663..320F}.  

The column density of H I along each sightline in the sample was
estimated by measuring Ly$\alpha$ absorption lines in the {\it
HST}/STIS spectra \citep{1994ApJ...427..274D}. This was done
simultaneously as part of the model fitting.  The estimates for N(H
I) are given in Table~\ref{tab_hi_results}. Also calculated and listed
in Table~\ref{tab_hi_results} are the 
gas-to-dust ratios, $N(H I)/E(B-V)$. There are also 21~cm observations
of H I for the four sightlines \citep{2011ApJ...726...39C}.  Unlike
the Ly$\alpha$ column densities which just measure the H I in front of
the star, the 21 cm column densities are for the entire sightline. So
the best estimate of the gas-to-dust ratio is 0.5 N(H I)/E(B-V) for
the 21 cm observations. These are also listed in
Table~\ref{tab_hi_results}. Because the 
fraction of the H I column density that is along the sightline to the
star in unknown, the uncertainties are assumed to be $\pm$100\%.

\section{Discussion}

The calculated extinction curves for the four new sightlines in M31
are shown in Figure~\ref{extcurves}. These curves have been corrected
for MW foreground so should reflect the extinction properties of
interstellar dust in M31. These curves are a great improvement on the
data presented in Figure~5 of \citet{1996ApJ...471..203B} which is an
average of several low-reddening sightlines. That curve seemed to show
a weak bump along with a MW-like FUV extinction.

The new curves sample dust in M31 at a range of galactocentric
distances and in different regions of M31 as shown in
Figure~\ref{m31image}. We are sampling sightlines separated by
kiloparsecs, 
much further apart than any sampled in our own Galaxy. The extinction
curve for J003944.71+402056.2 is the only sightline in the sample with
an estimated value of $R_V \sim 3.3$. This value is consistent with
its measured extinction curve shown in Figure~\ref{extcurves} which
looks very similar to the average MW (CCM $R_V=3.1$) extinction curve
overplotted in red.  This star appears to be associated with an H II
region, and shows strong diffuse interstellar band (DIB) features
\citep{1964ApJ...139.1027B,2011ApJ...726...39C}. J003944.71+402056.2
lies very close in projection to J003958.22+402329.0 but may not be
very close in three dimensions. The curve for J003958.22+402329.0 is
significantly different than J003944.71+402056.2 with a smaller FUV
curvature parameter ($c_4$) and a weaker bump as measured by
$c_3/\gamma^2$ \citep{1990ApJS...72..163F}. The extinction curve for
J004034.61+404326.1 is similar to J003958.22+402329.0 but with a
stronger bump. This sightline shows weak DIB features
\citep{2011ApJ...726...39C}.

\begin{figure*}[tbp]
\begin{center}
\epsscale{1.15}
\plotone{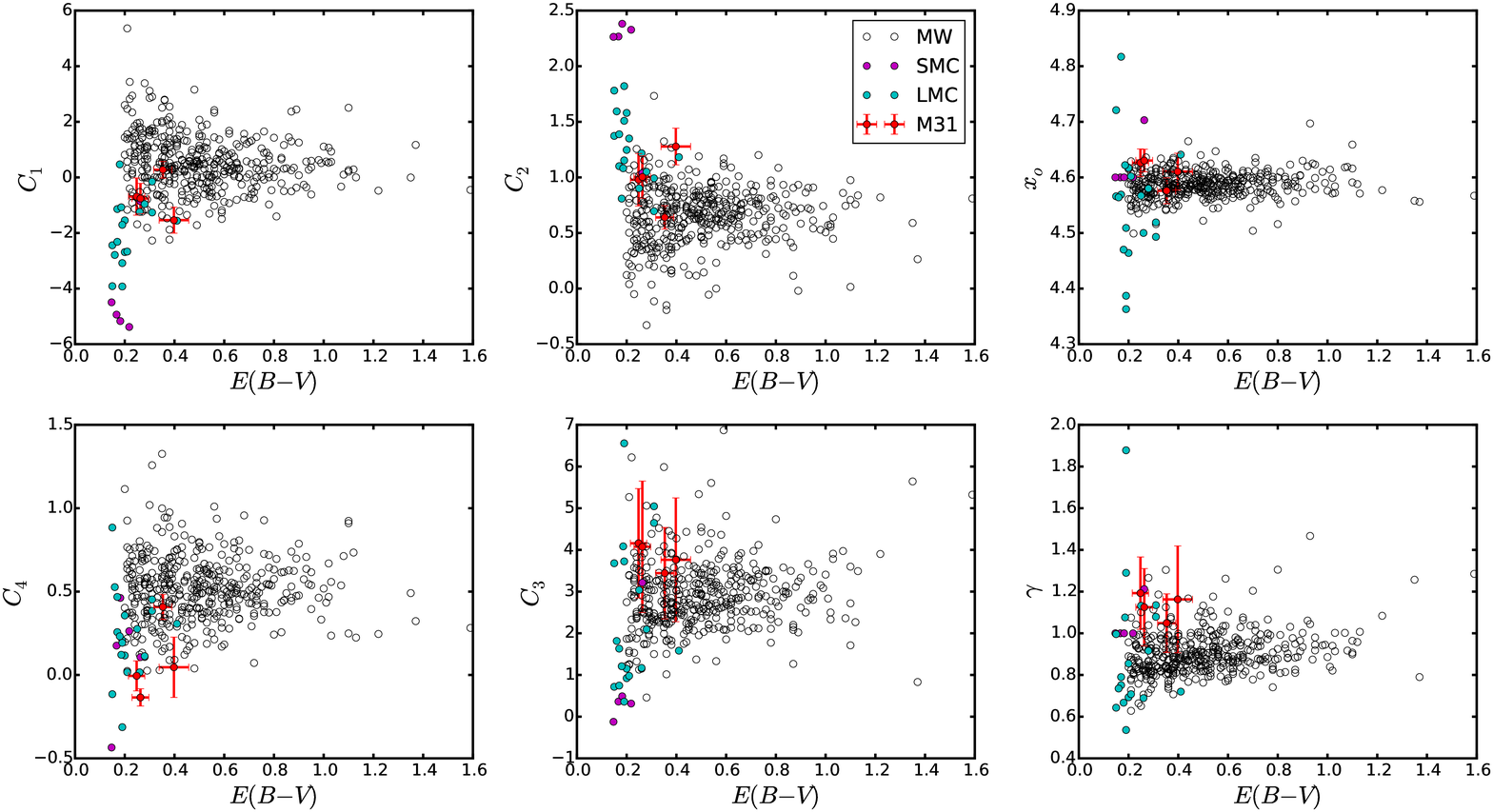}
\end{center}
\caption{The FM parameters plotted against ($E(B-V)$ for sightlines in
the MW, LMC, SMC, as well as the new sightlines in M31 presented here
\citep{2003ApJ...594..279G,2004ApJ...616..912V}.}
\label{fm}
\end{figure*}

The most reddened sightline, J004412.17+413324.2, shows a curve
reminiscent of the 30-Dor region of the LMC or a MW CCM $R_V \sim 2$
curve \citep{1999ApJ...515..128M}. But the lack of FUV curvature
($c_4$) makes it even flatter than the LMC-30-Dor curves. It is also
associated with an H II region
\citep{1964ApJ...139.1027B,2011ApJ...726...39C}. The
J004034.61+404326.1 and J003958.22+402329.0 extinction curves show a
similar wavelength dependence. This can be seen in Figure~\ref{fm}
where the FM parameters for MW, LMC, SMC and the new M31 sightlines
are plotted against $E(B-V)$. The Spitzer 8~\micron\ intensity is
relatively strong close to J004412.17+413324.2 possibly indicating the
presence of a dense dust cloud \citep{2011ApJ...726...39C}. It also
shows weak DIB features for its reddening as do sightlines in the LMC
\citep{2006A&A...447..991C}. The UV extinction observed through five
dusty clumps in the bulge of M31 has been measured using filter
photometry with {\it HST} and {\it Swift}
\citep{2014ApJ...785..136D}. They suggest that the curves are steeper
than the average MW extinction, perhaps with $R_V \sim 2.5$ but there
is a lot of scatter in their curves. Their result supports that of
\citet{2000MNRAS.312L..29M} which implied $R_V \sim 2.1$ using only
BVRI photometry for one sightline in the bulge of M31.  The extinction
curve for J004412.17+413324.2, which is the closest of the stars in
our sample to the M31 bulge, at a projected galactocentric distance of
5 kpc, resembles a MW curve with an $R_V \sim 2$ but is closer to an
LMC 30-Dor curve. These extinction measurements all show evidence for
low $R_V$, and small average grain sizes in the highest metallicity
regions of M31.

The gas-to-dust ratios measured here using Ly$\alpha$ absorption lines
for J003944.71+402056.2, J003958.22+402329.0, and J004034.61+404326.1
may be higher than the average MW value ($5.8 \times 10^{21}$ atoms
cm$^{-2}$ mag$^{-1}$) \citep{1978ApJ...224..132B} although these
values are quite uncertain. The projected galactocentric distance from
the center of M31 for these stars is between 8.8 and 14.0 kpc where
the metallicity is approximately Solar \citep{2012ApJ...758..133S}. 
J004412.17+413324.2, the most reddened
star, has the lowest N(H I) column density and therefore the lowest
gas-to-dust ratio in the sample, significantly lower than the average
MW value.  It is at a projected galactocentric distance of only 5.2 kpc which
corresponds to a metallicity about 1.6 Solar, so a low gas-to-dust
ratio would be expected \citep{2012ApJ...758..133S}. Unfortunately,
the measured Ly$\alpha$ N(H I) is very uncertain for this
sightline. But, the 21 cm data for the same sightline, listed in
Table~\ref{tab_hi_results}, also indicate that the gas-to-dust ratio
is low. The 21cm N(H I) estimates for the four stars imply lower
gas-to-dust ratios than the Ly$\alpha$ estimates, more in line with
the average MW value, but since the fraction of the N(H I) along the
line of sight to each star is uncertain, the column densities could be
higher or even lower. \citep{2011ApJ...726...39C}.  Taking an average
of the four sightlines using the 21 cm data, we get a gas-to-dust
ratio of $5.3 \times 10^{21}$ atoms cm$^{-2}$ mag$^{-1}$, close to
the MW value.

The measured extinction curves for the four M31 sightlines have been
used as inputs for Maximum Entropy Method (MEM) modeling of the
underlying dust grain populations.  This analysis is similar to that
applied previously to extinction curves in the MW and the Magellanic
clouds \citep{2003ApJ...588..871C}. 
We use a version of the MEM
extinction-fitting algorithm similar to that developed by \citet{1994ApJ...422..164K}. MEM uses the mass distribution
in which $m(a) da$ is the mass of dust grains per H
atom in the size interval from a to a + da rather than using the number of
grains as a constraint. 
A \citet{1977ApJ...217..425M}
(MRN-type) model becomes $m(a)$ $\propto$ a$^{-0.5}$.
The total mass of dust is constrained using both the gas-to-dust
ratio and the available abundances of iron, carbon and silicon.
The fraction of the available
silicon and carbon used in the MEM fits covers a very wide range. For
example, the MW, which has a relatively low gas-to-dust ratio and high
metal abundances, also requires almost 100\% of the available silicon
and and 80\% of the carbon \citep{2003ApJ...588..871C}. The LMC
(30-Dor) and the SMC Bar regions both have low metal abundances and
high gas-to-dust ratios. Both use 50\% or less of the available
silicon and 60-80\% of the carbon.  Three general factors determine
the fraction of silicon and carbon that any individual sight line will
use. First, the higher the gas-to-dust ratio is, the more metals are
available in the gas phase. Second, the higher the abundances of
metals are, the more material is available to make grains. Finally,
high values of the ratio of total-to-selective extinction, $R_V$,
imply a greater than average mass fraction in larger grains.

\begin{figure*}
\begin{center}
\epsscale{1.0}
\plottwo{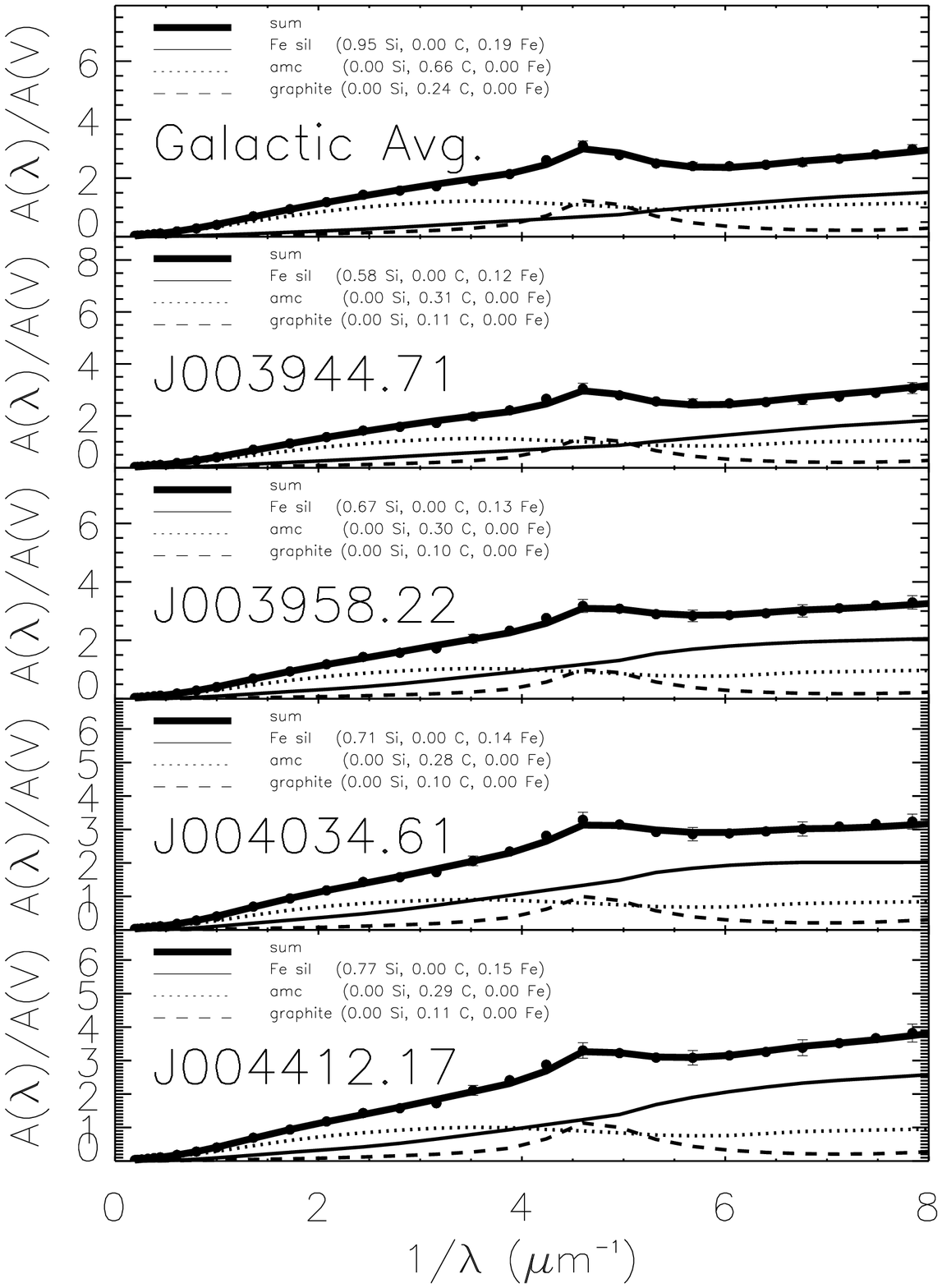}{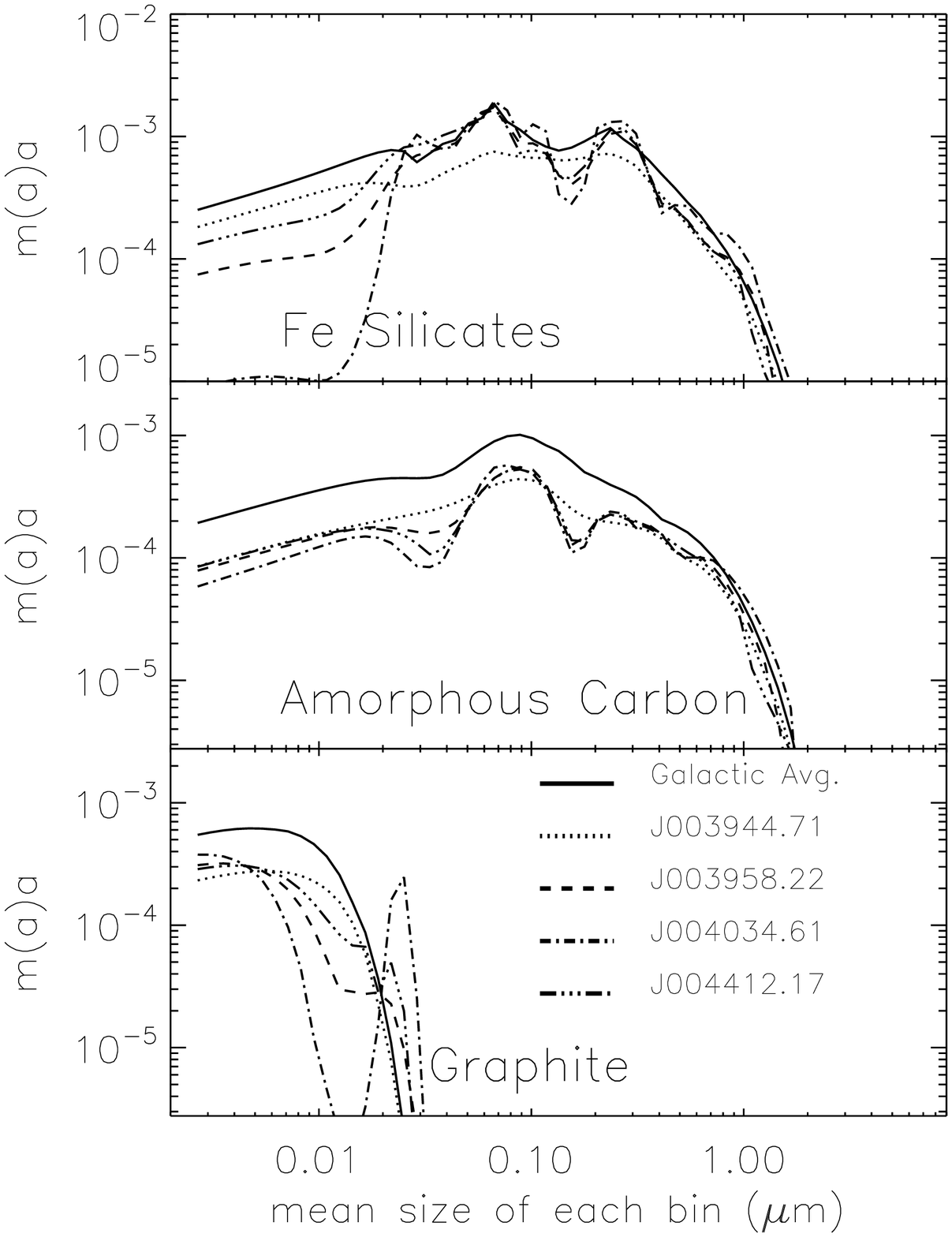}
\end{center}
\caption{Left: Three-component extinction models for the four M31
sight lines along with the Average MW extinction for comparison. Each
panel shows the model fit to the extinction curve, including the
contribution of each component. The fraction of the available Si, Fe,
and C (amorphous carbon and graphite) utilized is listed in the figure
legend.  Right: Three-component extinction models for the same sight
lines. Each panel contains the resulting mass distributions relative
to the mass of hydrogen.}
\label{mem}
\end{figure*}

The MEM fit
results are shown in Figure~\ref{mem}. The fits were made using three
interstellar dust components, Fe-silicate, graphite, and amorphous
carbon
\citep{1994A&A...292..641J,1995A&A...300..503D,2003ApJ...588..871C}. We
have adopted the following MW abundances for the MEM fitting, Si/H =
$4.0\times 10^{-5}$, C/H = $3.20\times 10^{-4}$, and Fe/H =
$4.1\times 10^{-5}$ \citep{2003ApJ...588..871C}.  As described above,
the measured gas-to-dust ratios may be higher than the MW for three of
the sightlines and lower for J004412.17+413324.2 but have large
uncertainties. Also, based on their projected positions in M31, the
abundances should be near Solar for J003944.71+402056.2,
J003958.22+402329.0, and J004034.61+404326.1 and super-Solar for
J004412.17+413324.2. The latter sightline has the lowest gas-to-dust
ratio as expected. Therefore, three MEM fitting runs were performed
for the four M31 sightlines, assuming Solar abundances along with
three different gas-to-dust ratios, MW ($5.8 \times 10^{21}$ atoms
cm$^{-2}$ mag$^{-1}$), 1/2$\times$MW, and 2$\times$MW. The results for
the run with Solar abundances and the 2$\times$MW gas-to-dust ratio
are shown in Figure \ref{mem}. Also shown in Figure \ref{mem} are the
values for the MW CCM $R_V=3.1$ average extinction curve.

Of the three MEM runs, in the first two, with Solar abundances, and MW
and 1/2$\times$MW gas-to-dust ratios, there is not enough Si and C
available to make the dust grains required to fit the observed
extinction. In the last case, shown in Figure \ref{mem}, the available
abundances are enough to make the grains needed. The four M31
sightlines modeled here use 53-74\% of the Si, 11-15\% of the Fe,
13-16\% of the graphite, and 38-41\% of the amorphous carbon
available. The results show that the M31 extinction curves can only be
fit where the abundances are greater than Solar, the gas-to-dust ratio
is greater than the average MW ratio or a combination of both. The
large measured gas-to-dust ratios for J003944.71+402056.2,
J003958.22+402329.0, and J004034.61+404326.1 and Solar abundances are
needed to fit those sightlines. The most reddened sightline,
J004412.17+413324.2, which lies closest to the center of M31 is
problematic. The abundances for this sightline would need to be very
super-Solar if the small measured gas-to-dust ratio is
correct. Unfortunately, there is not much information on $R_V$ for
these sightlines, only J003944.71+402056.2 with $R_V=3.3$ has an
estimate which seems reasonable for its observed extinction curve.

\clearpage

\section{Summary} 

The new UV extinction curves presented here provide
a tantalizing glimpse into the interstellar dust properties in
M31. Unlike the MW, the dust in M31 can be sampled across the whole
galactic disk. These results are a significant improvement on previous
M31 extinction curves \citep{1996ApJ...471..203B}. The sample is small
but the extinction properties range from a curve very similar to CCM
$R_V=3.1$ seen toward J003944.71+402056.2 which is at a projected
distance from the galactic center of 14 kpc to a curve similar to LMC
(30-Dor) extinction seen toward J004412.17+413324.2 which is at a
projected distance of 5 kpc.

The new extinction curves show similarities to those seen the MW and
the LMC. The highest metallicity sightline which is closest to the M31
bulge, and has the lowest gas-to-dust ratio shows an extinction curve
consistent a low value of $R_V$ or possibly with the LMC 30-Dor
region. Many more sightlines need to be studied to map out the
extinction properties of M31 and to investigate the link between UV
extinction properties and global characteristics such as metallicity
and star formation activity.

This study was supported by grant HST-GO-12562.01 from
NASA/STScI. Some of the data presented in this paper were obtained
from the Mikulski Archive for Space Telescopes (MAST). STScI is
operated by the Association of Universities for Research in Astronomy,
Inc., under NASA contract NAS5-26555. We would like to thank Jonathan
Sick for providing JK photometry for this study.

\bibliography{/Users/gclayton/projects/latexstuff/everything2}

\end{document}